# PRINCIPLES FOR DIGITAL PRESERVATION

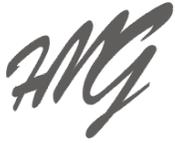

H.M. Gladney

**HMG Consulting**
**Saratoga, CA 95070**

© 2004, H.M. Gladney

**10 November 2004**
**revised 15 December 2004**

**The immense investments in creating and disseminating digitally represented information have not been accompanied by commensurate effort to ensure the longevity of information of permanent interest. Asserted difficulties with long-term digital preservation prove to be largely underestimation of what technology can provide. We show how to clarify prominent misunderstandings and sketch a 'Trustworthy Digital Object (TDO)' method that solves all the published technical challenges.**

Most information is now "born digital". Much is disseminated only in digital form. However little of this is provided in forms that ensure its perpetual intelligibility or that include evidence that it can be trusted for sensitive applications.

Most articles about digital preservation come from the cultural heritage community, which expresses distress that the information technology community is not involved. The NDIIPP (National Digital Information Infrastructure Preservation Plan) [LC] expresses urgency for preserving authentic digital works. However, since the 1995 appearance of *Preserving Digital Information* [Garrett], little progress has been made towards technology for reliable preservation of substantial collections. [Marcum], [Waters]

Most of the preservation literature draws its examples from scholars' and artists' interests. We anticipate that the needs expressed will expand to those of businesses wanting safeguards against diverse frauds, attorneys arguing cases based on the probative value of digital documents, and our own dependencies on personal medical records.

The current article deals exclusively with challenges created by technological obsolescence and the demise of information providers. [Thibodeau] summarizes preservation know-how with:

- proven methods for preserving and providing sustained access to electronic records were limited to the simplest forms of digital objects;
- even in those areas, proven methods were incapable of being scaled [for] the expected growth of electronic records; and
- archival science had not responded to the challenge of electronic records sufficiently to provide a sound intellectual foundation for articulating archival policies, strategies, and standards for electronic records.

We summarize a design[1] that addresses all technical issues reported in the preservation literature.

---

[1]    Detailed designs are being published in the ACM Transactions on Information Systems.

© 2004, H.M. Gladney

## What Would a Preservation Solution Provide?

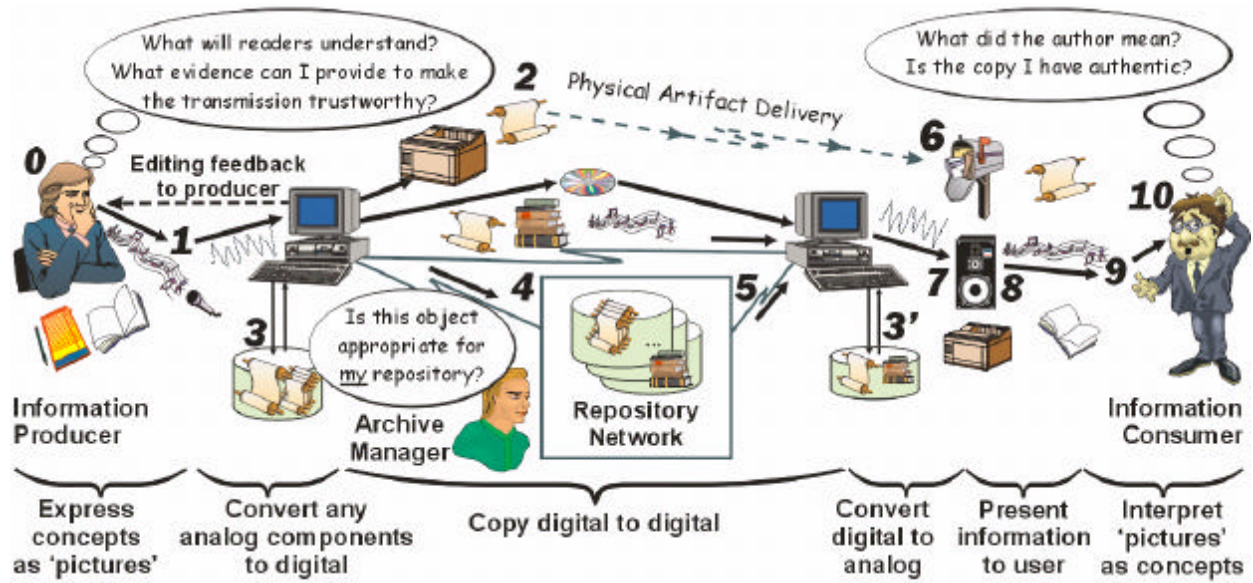

**Figure 1: Documentary information interchange and repositories**
(The object numbering is taken from [Gladney 2].)

What might someone a century from now want of information stored today?  Figure 1 helps us discuss preservation reliability.  In addition to what content management offerings[2] and published metadata schema[3] already provide, a complete solution would:

- Ensure that a copy of every preserved document survives as long as wanted;

- Ensure that authorized consumers can find and use any preserved document as its producers intended, avoiding errors introduced by third parties;

- Ensure that any consumer can decide whether information received is sufficiently trustworthy for his application; and

- Hide technical complexity from end users (both information producers and consumers).

Viable solutions will allow repositories and their clients to use deployed content management software without disruption.

# CHALLENGES EXPOSED BY PRIOR WORK

Information in physical books, on other paper media, and in other analog forms cannot be copied without error and invariably contains accidental information that digital representations can avoid.  Perfect digital copying is possible, and both contributes to the challenge of preserving digital content and to its solution.  Preservation can be viewed as a special case of information interchange—special because information consumers cannot obtain information producers' responses for puzzling aspects or missing information.

## Pervasive Focus on Repositories

Most preservation literature focuses on "Trusted Digital Repositories".  [Beagrie] calls for certifications that might lead to public announcement that an institution has correctly executed sound preservation practices. However, to execute partly human procedures faithfully over decades would be difficult and expensive. Repository-centric proposals betray problems that call the direction into question:

- They depend on an unexpressed premise—that exposing an archive's procedures can persuade its clients that its content deliveries will be authentic. Such procedures have not yet been described, much less justified as achieving what their proponents apparently assume.

- Audits of a digital archive—no matter how frequent—cannot prove that its contents have not been improperly altered by employees or hackers many years before a sensitive document is accessed.

- The new code needed for digital preservation is likely to be mostly workstation software, not server software, so that the people focusing on repositories find it difficult to design solutions.

---

[2] Content management is not discussed in the current article because available software is sufficient for archives with at most modest and obvious extensions.

[3] These include general schema proposed for standardization, such as METS sponsored by the Library of Congress, and many topic- or discipline-specific extensions.  A WWW search for "metadata schema" yielded ~29,000 hits.

As an objective, 'trusted' is misleading. Instead, one should focus on encapsulating digital objects so that they are **trustworthy**.

## What's 'the Original'? What's 'Authentic'?

In casual conversation, we often say that the copy of a recording is authentic if it closely resembles the original. But consider, for example, an orchestral performance, with sound reflected from walls entering imperfect microphones, signal changes in electronic recording circuits, and so on, until we finally hear a television rendering. Which of many different signal versions is 'the original'?

Difficulties with 'original' and 'authentic' are conceptual. Nobody creates an artifact in an indivisible act. What people consider to be an original or a valuable derivative version is somebody's subjective choice, or an objective choice guided by subjective social rules. We can describe any version objectively with provenance metadata that express everything important about the creation history.

Conventional definitions, such as "authentic: of undisputed origin; genuine." [Concise Oxford English Dictionary], do not help operationally. For signals, for material artifacts, and even for natural entities, the following definition captures what people mean when they say 'authentic'.

> **Given a derivation statement R,**      **"V is a copy of Y ( V=C(Y) )",**
>      **a provenance statement S,**      **"X said or created Y as part of event Z", and**
>      **a copy function,**      **"C(y) = $T_n(\ldots (T_2( T_1(y) )))$,"**
> **we say that V is a *derivative* of Y if V is related to Y according to R.**
> **We say that "by X as part of event Z" is a *true provenance* of V if R and S are true.**
> **We say that V is *sufficiently faithful* to Y if C conforms to social conventions for the genre and for the circumstances at hand.**
> **We say that V is an *authentic copy* of Y if it is a *sufficiently faithful derivative* with *true provenance*.**

Each $T_k$ represents a transformation that is part of a Figure 1 transmission step. To preserve authenticity, the metadata accompanying the input in each transmission step should be extended by including a $T_k$ description. These metadata might identify who is responsible for each $T_k$ choice and other circumstances important to consumers' judgments of authenticity.

## Preserving Dynamic Behavior

A prominent collaborative archivists' project suggests conceptual difficulty with preserving "dynamic objects" (representations of artistic and other performances) digitally. [Duranti] We see no new or difficult technical problem. What differs for different object types is merely the ease of changing them.

A repeat R(t) of an earlier performance P(t) would be called authentic if it were a faithful copy except for a constant time-shift, $t_{start}$, i.e., if $R(t)=P(t-t_{start})$. This seems simple enough and capable of describing any kind of performance. Its meaning is simpler for digital documents than for analog recordings because digital files already reflect the sampling errors of recording performances that are continuous in time.

The archivists expressing difficulty with dynamic digital objects do not express similar uncertainty about analog recordings of music.

# TRUSTWORTHY DIGITAL OBJECT (TDO) METHODOLOGY

TDO methodology focuses on preservation objects and defines methods for making their authenticity reliably testable and for assuring that eventual users will be able to render or otherwise use their contents. Our objectives suggest solution components that can be nearly independently addressed:

I. Content servers that store packaged works, and that provide search and access services.[2]

II. Replication mechanisms that protect against the loss of the last remaining copy of any work. [Reich]

III. A method for packaging a work together with metadata that includes provenance assertion and reliable linking of related works, ontologies, rendering software, and package pieces with one another.

IV. Standard bibliographic metadata and topic-specific ontologies defined, standardized, and maintained by professional communities.[3]

V. A bit-string encoding scheme to represent each content piece in language insensitive to irrelevant and ephemeral aspects of its current computer environment.

The TDO base is record schema for long-term holdings. (Figure 2) To prepare the object set that makes up a work, an editor causes conversion of each content bit-string into a durably intelligible representation and collects the results, together with standardized metadata, to become the payload of a new TDO. In addition to its payload, each TDO has a protection block into which a human editor loads metadata and records relationships among parts of the new TDO, and between it and other objects. The final construction step, executed at a human agent's command, is to seal all these pieces in a single bit-string with a *message authentication code*. In a valid TDO representing some version of an object:

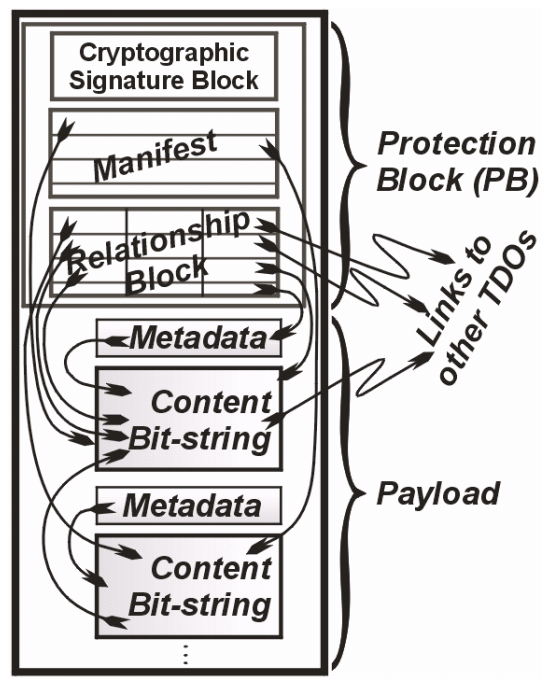

**Figure 2: A Trustworthy Digital Object (TDO)**

- The bit-string set that represents the version is XML-packaged with registered schema.

- These bit-strings and metadata are encoded to be platform-independent and durably intelligible.

- The metadata include identifiers for the version and for the set of versions of the work.

- The package includes or links reliably to all metadata needed for interpretation and as evidence.

- All these contents are packaged as a single bit-string sealed using cryptographic certificates based on public key message authentication.

- Each cryptographic certificate is authenticated by a recursive certificate chain.

In ancient times, wax seals impressed with signet rings were affixed to documents as evidence of their authenticity. A digital counterpart is a message authentication code firmly bound to each important document. **Evidence After Every Witness is Dead** [Gladney 3] describes the structure and use of each TDO, emphasizing the metadata portions suggested by Figure 2. The design includes the following features.

(i) Each TDO contains its own world-wide eternal and unique identifier and its own provenance metadata, and is cryptographically sealed to prevent undiscoverable changes;

(ii) References to external objects are accompanied by their referents' message authentication codes;

(iii) Certification keys are themselves certified. This recursion is grounded in the published and annually changed public keys of institutions that people trust to be honest witnesses. Sealing of a preserved document by such an institution creates durable evidence of its publication date;

(iv)   Each human being who edits a work being prepared for archival deposit nests or links the version he started with, thereby creating a reliable history;

(v)    Each participant in the creation sequence usually is, or readily can become, acquainted with his predecessor and his successor.  Thus the public keys that validate authorized version deliveries can readily be shared without depending on a Public Key Infrastructure (PKI) certificate authority.  This arrangement avoids well-known PKI security risks.

Relatively simple data formats can be saved with ad hoc methods that ensure their eventual intelligibility. For other data formats, ***Durable Encoding for When It's Too Late to Ask*** [Gladney 2] teaches how to encode any kind of Figure 2 content bit-string to be durably intelligible or useful.  Its features include:

(vi)   That we enable each information producer to separate irrelevant information, such as operating system details, from information essential to his intentions, encoding only what's essential;

(vii)  Rewrite to the code of a Turing-complete virtual machine (extended to handle concurrency and real-time services)—an application of the Church-Turing thesis that any program or rule set producing a finite sequence can be implemented by a simple machine.

(viii) And that such machines can themselves be described completely and unambiguously.

A producer typically tries to encode information so that each consumer can read or otherwise use the content.  In an ideal Figure 1 scenario, perfection would be characterized by the consumer understanding exactly what the producer intended to communicate.  However, in addition to the consequences of human imperfections of authors and editors, the **0➔1** and **9➔10** steps suffer from unavoidable language limitations.  (Jargon, expectations, world views, and ontologies are at best imperfectly shared.)  *I cannot tell you what I mean.  I cannot know how you interpret what I say*.

Such difficulties originate in the theoretical limits of what machines can do.  How we might manage them will be discussed in two further articles.  ***Syntax and Semantics—Tension between Facts and Values*** [Gladney 4] provides philosophical arguments that the methods described in the prior papers do as much as mechanical methods can do towards preserving digital information, and that these methods attempt no more.  A second work in progress examines what information producers can do to minimize eventual consumers' misinterpretations, given that communication invariably confounds intentional with accidental information.

# DISCUSSION

Premature digital preservation deployment would risk that flaws might not be discovered until after large investments into creating archival holdings.  Errors might distort meanings (for texts) or behaviors (for programs).  The questions reach into epistemology—the philosophical theory of what can be objectively known and reliably communicated, in contrast to what must forever remain subjective questions of belief or taste.  We are therefore reluctant to implement pilot installations until we have considered the applicable philosophy thoroughly and until experts have had opportunity to criticize TDO design.

## What's Missing from the U.S. Digital Preservation Plan?

Engineers want questions that can be answered objectively.  They expect plans to be clear enough so that every participant and every qualified observer can understand what work is committed and can judge whether progress is being achieved.

We expect a plan to articulate each objective concisely, the resources needed to meet it, commitments to specific actions, a schedule for each delivery, and a prescription for measuring outcomes and quality.  If the plan is for a large project, we expect it to be expressed in sections that separate teams can address relatively independently.  If the resources currently available are inadequate, we expect the plan to identify each shortfall.  Finally, if a team has already worked on the topic, we expect its plan to list its prior achievements.

NDIIPP funding is commensurate with that for all foreign preservation work combined.  Regrettably, the technical portions of [LC] contain little more than vague generalities and decade-old ideas.  It identifies few technical specifics, no target dates, and few objective success measures.  Engineers will find little to work with.  This is troubling for an initiative launched almost four years ago.

# Competitive Evaluation

Firm assertions of TDO packaging advantages over alternatives would be premature before we have deployed a complete pilot. Ideally, we would compare our design to alternatives. However, nobody has designed one. Notwithstanding such uncertainties, we believe that, in addition to satisfying our starting objectives, TDO packages will exhibit the following desirable characteristics.

(1)  Consumers will be able to evaluate TDO content authenticity without help from administrators.

(2)  Metadata-to-object dissociation will occur at most rarely, and will be discernible when it happens.

(3)  Correct information delivery will be insensitive to Internet security risks. Objects might disappear, but if a TDO is delivered, its integrity can be validated.

(4)  Identifier naming authorities are not needed, thereby avoiding overhead. Specialized name-to-location resolvers might not be needed, because popular Web crawlers can readily include the function. A TDO can be replicated across the Internet without explicitly managing resolver updates.

(5)  Collection management can be simplified by exploiting TDO link reliability. If metadata are sufficiently standardized, users will be able to use automatic tools to create personal digital library catalogs that suit their special needs and preferences.

(6)  TDO software can be brought into service without disrupting installed digital libraries. Preservation objects can be stored, catalogued, and served by any of several content manager offerings.

What will make implementations easy to tailor is that good tools exist for XML. What will make them scalable is that TDO structure is recursive and uses links extensively.

## Conclusions

Most preservation literature emphasizes the perspectives of archiving institutions. This and the supporting TDO reports focus on end users' needs because these have precedence over repository needs. We have articulated principles for a TDO design that addresses every technical problem and requirement articulated in the literature. Its central elements are an encapsulation scheme for digital preservation objects and encoding using extended Turing-complete virtual machines. Correct TDO implementations will allow preservation of any type of digital information and will be as efficient as any competing solution.

We request searching critical examination of the work by readers. Public discussion is called for because "getting it right" is too important for anything short of complete transparency.

## Acknowledgements

John Bennett, Peter Farwell, Tom Gladney, Raymond Lorie, Peter Lucas, and John Swinden provided detailed discussions of TDO methodology.

## Bibliography

[Beagrie]      Beagrie, Neil. et al. *Trusted Digital Repositories: Attributes and Responsibilities,* RLG-OCLC Report, 2002. http://www.rlg.org/longterm/repositories.pdf

[Garrett]      Garrett, John. et al. *Preserving Digital Information: Report of the Task Force on Archiving of Digital Information*, Commission on Preservation and Access and The Research Libraries Group, 1995.

[Duranti]      Duranti, Luciana. *The Long-term Preservation of the Dynamic and Interactive Records of the Arts, Sciences and E-Government*, Documents Numerique 8(1), 1-14, 2004.

[Gladney 2]    Gladney, H.M. Lorie, Raymond. *Trustworthy 100-Year Digital Objects: Durable Encoding for When It's Too Late to Ask,* 2003, in review by ACM Transactions on Information Systems.

[Gladney 3]    Gladney, H.M. *Trustworthy 100-Year Digital Objects: Evidence After Every Witness is Dead*, ACM Trans. Info. Sys. 22(3), 406-436, July 2004.

[Gladney 4]   Gladney, H.M. *Trustworthy 100-Year Digital Objects: Syntax and Semantics—Tension between Facts and Values,* 2004 preprint at http://eprints.erpanet.org/archive/00000051/.

[LC]   Library of Congress, *Preserving Our Digital Heritage: Plan for the National Digital Information Infrastructure and Preservation Program,* 2003. http://www.digitalpreservation.gov/repor/ndiipp_plan.pdf

[Marcum]   Deanna B. Marcum, *Research Questions for the Digital Era Library*, Library Trends 51(4), 636-651, Spring 2003.

[Reich]   Reich, Vicky. Rosenthal, David S.H. *LOCKSS: A Permanent Web Publishing and Access System*, D-Lib Magazine 7(6), June 2001.

[Thibodeau]   Thibodeau, Kenneth. *Knowledge and action for digital preservation: Progress in the US Government*, Proc. DLM-Forum 2002, 175-9, 2002.

[Waters]   Waters, Donald. *Good Archives Make Good Scholars: Reflections on Recent Steps Toward the Archiving of Digital Information*, in Council on Library and Information Resources pub107, 2002. Other essays in these proceedings help portray the 2002 level of understanding.